# Reexamining Statistical Significance and P-Values in Nursing Research: Historical Context and Guidance for Interpretation, Alternatives, and Reporting

Christopher Holmberg [1,2]

1. Institute of Health and Care Sciences, University of Gothenburg, Gothenburg, Sweden

2. Department of Psychotic Disorders, Sahlgrenska University Hospital, Gothenburg, Sweden

ABSTRACT
Nursing should rely on the best evidence, but nurses often struggle with statistics, impeding research integration into clinical practice. Statistical significance, a key concept in classical statistics, and its primary metric, the p-value, are frequently misused. This topic has been debated in many disciplines but rarely in nursing.
The aim is to present key arguments in the debate surrounding the misuse of p-values, discuss their relevance to nursing, and offer recommendations to address them.
The literature indicates that the concept of probability in classical statistics is not easily understood, leading to misinterpretations of statistical significance. A substantial portion of the critique concerning p-values arises from such misunderstandings and imprecise terminology. Consequently, some scholars have argued for the complete abandonment of p-values.
Instead of discarding p-values, this article provides a more comprehensive account of their historical context and the information they convey. This will clarify why they are widely used yet often misunderstood. Additionally, the article offers recommendations for accurate interpretation of statistical significance by incorporating other key metrics. To mitigate publication bias resulting from p-value misuse, pre-registering the analysis plan is recommended. The article also explores alternative approaches, particularly Bayes factors, as they may resolve several of these issues.
P-values serve a purpose in nursing research as an initial safeguard against the influence of randomness. Much criticism directed towards p-values arises from misunderstandings and inaccurate terminology. Several considerations and measures are recommended, some which go beyond the conventional, to obtain accurate p-values and to better understand statistical significance. Nurse educators and researchers should considerer these in their educational and research reporting practices.

# 1. Introduction

Scholarly efforts to correctly understand p-values and statistical significance are long overdue. [1,2] Given the widespread misuse of those concepts, some scholars have even proposed altogether discarding significance testing and p-values from research. [3] However, such a ban is neither feasible nor preferable. [4] After all, in nursing, researchers, clinicians, and patients need to know the probability associated with the hypothesis of the efficacy of an intervention, such as a health promotion intervention or psychosocial intervention [5] Even so, p-values remain widely misunderstood in many scientific disciplines, including nursing. [1,6]

Nurses' understanding and knowledge formation, derived from scientific studies, contribute to the establishment of a nursing practice which is more grounded in evidence. Accurate inference and utilization of studies using statistics are thus important competencies for nurses in delivering high-quality patient care. [7] As nursing undergoes a generational transformation involving the widespread application of electronic health records, nurses engaged in research need to be well-versed in a comprehensive set of quantitative methods [8] and need skills in using statistics in order to advance the unique discipline of nursing. Because nurses seek to provide evidence-based, cost-effective, high-quality care, it has become increasingly common for them to conduct experimental research to test and evaluate nursing interventions by performing, for example. randomized controlled trials. [9] Given those trends, education for nurses in using statistics should begin at the undergraduate level. Although most undergraduate nursing students may not be involved in designing research projects or conducting statistical analyses, they need to know basic concepts of statistics in order to develop critical thinking skills. [10]

Although a more thorough understanding of p-values and significance testing is warranted in nursing, several studies have identified that knowledge of and attitudes toward statistics are weaker among nurses than other health care professionals. [11,12] For nurses, difficulty in understanding statistics ranks among the most common barriers to implementing research findings in clinical practice. [13,14] The inadequate use of statistics among nurses has also manifested in advanced nursing degree projects. [15] Regarding published work, a bibliometric analysis of 30 nursing journals revealed that more than 90% (28/30) of the journals included articles containing misrepresented p-values. [16] Beyond that, the use of significance testing has been criticized in many disciplines, perhaps most prominently in psychology, in which it has been spotlighted as a primary contributor to the replication crisis. [17] Nevertheless, nursing scholars have largely been absent from the scientific debate over controversies surrounding the use of statistical significance and p-values. [6,18,19] And those articles that have addressed these issues have normally not included alternative approaches, [20,21] such as Bayes factors, or covered overarching aspects like data management, which are essential for mitigating publication bias stemming from p-value misuse. In response, this article aims to clarify key misunderstandings within the debate, place them in the context of nursing, and offer recommendations, some which go beyond the conventional, to promote more robust practices.

## 2. Introduction: Definition and historical overview

Simply put, the p-value represents the probability in a specific model that a statistical metric of the data (e.g., mean differences between groups in a sample) is equal to, or more extreme than, the observed value. [1] P-values are thus conditional probabilities calculated under the assumption that the model (the null) is true. P-values are therefore often used to determine whether sample estimates differ significantly from hypothesized values. Although p-values can thus be viewed purely as a mathematical product, they rely on a specific understanding of probability. In fact, the concept of probability is central in statistics, for most statistical approaches and inferences are underscored by measure-theoretic probability theory. [22]

In this article, probability is understood as being physical, insofar as it reflects the occurrence of a situation, often described as chance. Physical probability is theoretically related to classical (i.e., frequentist) statistics, meaning that probability is interpreted in view of the long-term frequency of repeatable studies. An alternative way of conceiving probability highlights the epistemic function that it plays in formulating hypotheses—in that understanding, the degree of belief, or degree of support, that a situation will occur. That view on probability undergirds another perspective in statistics: Bayesian statistics. [23] Although that approach is not the article's focus, it is discussed considering the Bayes factor, presented as an alternative to p-values.

However, understanding why p-values rank among the most ubiquitous but misunderstood statistical metrics [24,25] requires first understanding the history of their development. Thus, in what follows, the key approaches are briefly introduced chronologically, beginning with Fisher's tests of significance and Neyman and Pearson's tests of statistical hypotheses [26] developed in the 1920s. After that, a combination of the two types of testing, namely null hypothesis significance testing, first elaborated by Lindquist [27], is presented.

### 2.1. Fisher, Neyman and Pearson, and Null Hypothesis Significance Testing

Although the p-value was first introduced in 1900 by Pearson with the chi-square test [28], it was Fisher who, in 1925, first articulated approaches to assess p-values in a wide variety of circumstances. Whereas some steps in Fisher's method are settled a priori (e.g., in formulating hypotheses), the method is predominantly inferential. Neyman and Pearson, seeking to improve Fisher's approach, adjusted it to focus more on decision-making between competing hypotheses.[26] Their method, usually applied at the a priori stage of planning studies, entails more calculations and estimations than Fisher's [25,29], and their mathematically inclined model introduced some key statistical terms (e.g., Type 1 error). The comparison of Fisher's approach and Neyman–Pearson's approach is inspired by the work of Halpin and Stam, Lehmann, and Perezgonzalez. [29–31]

| Fisher's approach | Neyman–Pearson's approach |
|---|---|
| *Before the test* | *Before the test* |
| • Select a test based on the research objective and how the variables have been measured. | • Estimate the effect size in the population and select an optimal test based on its power; parametric tests are preferred over non-parametric ones if they meet the assumptions. |
| • Derive the null hypothesis (H0) from the test as a precise statistical hypothesis. Certain parameters should be estimated from the sample, including | |
| degrees of freedom, whereas others (e.g., frequency distributions under a specific distribution) can be inferred.<br>• Calculate the probability (i.e., p-value) of the results given H0, by using the defined analogous distribution. In the null distribution, data closer to the mean have a higher likelihood of occurrence. | • Formulate the primary hypothesis and two or more competing hypotheses. The most central hypothesis for the study (i.e., which should not be rejected too often) is used; Type I errors occur when the primary hypothesis is incorrectly rejected. Therein, alpha ($\alpha$) represents the probability that Type I errors will occur over time.<br>• Establish an alternative hypothesis and consider Type II errors (i.e., the primary hypothesis is wrongly retained). The probability of Type II errors' occurring over time is represented by beta ($\beta$).<br>• Gauge the sample size needed for adequate power ($1 - \beta$). Power is the opposite of Type II error (i.e., $1 - \beta$).<br>• Using some of the above parameters (e.g., $\alpha$ and N), assess the critical value of the test for deciding between hypotheses. |
| *Following the test* | *Following the test* |
| • Evaluate the results' statistical significance to detect ones of particular interest (i.e., with a low probability of occurring). In that case, the p-value is testimony against the null hypothesis, and smaller p-values provide stronger proof against the occurrence of results due to random variation.<br>• Interpret statistically significant findings in one of two ways: as rare findings that occur only with a certain probability, p, or lower or as indicating that the null hypothesis does not fully explain the results. | • Compute the test value by estimating some unknown population parameters from the sample (e.g., variance) while deriving other parameters (e.g., frequency distributions) under a certain distribution.<br>• Decide between hypotheses using the prior specifications as follows: (a) Reject the primary hypothesis if observations were within the critical region; (b) accept the primary hypothesis if observations are outside the critical region and if the test has sufficient power; and (c) conclude nothing if observations appear outside the critical region and the test has insufficient power. |

Fisher's approach offers certain advantages. For one, it is flexible, because most of the work is done a posteriori, and thereby allows using different tests and testing null hypotheses. For another, it is suitable for ad hoc research projects and exploratory research. [32] A limitation, however, is that the approach is largely inferential—it extends findings from the sample to the population under consideration—and thus delimited to populations that can be assumed to present parameters like the sample estimates. Beyond that, it neither considers power analysis nor requires explicitly stating alternative hypotheses to be evaluated. Of course, Neyman–Pearson's approach also offers advantages, including that it is more powerful than Fisher's for testing data over time and thus more suitable for repeated sampling and measures. Among the approach's drawbacks, however, is its rigidity in requiring many a priori steps and, in turn, its inflexibility relative to Fisher's in accommodating tests not considered in advance.

Beyond those differences, however, the numerous similarities between the approaches can easily lead to their conflation, as demonstrated by null hypothesis significance testing, a sort of combination of the approaches. [33] However, many statisticians and textbooks [25] seem to be unaware that the p-values and significance testing conceived by Fisher are paradigmatically dissimilar to Neyman–Pearson's hypothesis testing model. [30] For p-values in particular, such confusion often stems from a misconception of statistical versus real-world significance and the testing of statistical significance without testing clear hypotheses. That confusion has led some nursing scholars to propose that significance testing should be limited in nursing education, following the idea that most nurses are simply not statisticians. [34]

Although reasons for criticizing traditional approaches to testing significance abound, many of which overly focus on specific p-values for decision-making, such arguments provide no firm basis for blaming p-values for behaving as intended or for their misuse or misinterpretation. Such criticism typically comes from pedagogues and researchers who fail both to grasp the inferential meaning of p-values and to describe hypothesis testing appropriately for practical needs. [35] Assuming that testing significance and using p-values will continue to be commonplace in nursing research, proposing remedies for common forms of their misuse is important. Among others, constructs in Neyman–Pearson's approach, particularly Type I and II errors, and associated constructs such as effect size can benefit null hypothesis significance testing and the use of p-values. [29,36] That and other recommendations, shown in Figure 1, are described in the following sections.

**Statistical significance testing in nursing education and research: Recommendations for more rigorous practices**

| Presenting and applying p-values | Communicating statistical significance | Using additional key metrics | Aspects of organizational and data management |
|---|---|---|---|
| • Present exact p-values<br>• Lower the alpha (α) level in certain cases | • Use correct terminology<br>• Do not confuse statistical with clinical significance or importance | • Always report confidence intervals and effect sizes<br>• Consider the Bayes factor | • Pre-register studies when relevant<br>• Enable data availability |

**Figure 1.** Recommendations for more robust statistical significance testing in nursing education and research.

## 3. Presenting and applying p-values
Because I argue that p-values do provide meaningful information, the focus of the following recommendations is to present and apply them accurately.

### 3.1. Present exact p-values
As mentioned, Fisher [37] conceived the p-value as an index for measuring the divergence between collected data and the population in question—or else a null hypothesis. Along those lines, a p-value can be understood as a measure of the strength of an indication against the null hypothesis as observed in the sample data. As such, a smaller p-value suggests a stronger indication.

However, Fisher would argue that simply identifying a result or effect of a parameter of interest (e.g., difference between interventions) as being significant if p < .05 and not significant otherwise represents a narrow understanding of statistics. For one, the .05 cutoff is arbitrary, and it does not consider elements of the research setting or study design. [38] For another, amid the debate over best practices in statistical inference, a consensus has emerged that the often routine practice of relying solely on p-values to dichotomize the results of single studies—that is, treating a result with a p-value below a certain threshold as being significant and otherwise not significant—should be abandoned. [39] Instead, the preferred alternative is to view p-values as continuous in that they provide a relative measure of an indication's strength. [40]

To facilitate that approach, authors should provide computed p-values so that readers can interpret the results as they deem appropriate. For example, although a p-value of .052 is nearly the same as one of .048, the first value has typically been disregarded; though that value may be clinically significant, the study in which it was calculated contained elements that negatively affected the p-value (e.g., small sample size). By contrast, when p-values are viewed as being continuous, they can be interpreted against each other. Considering relatively small p-values (e.g., p < .10) as 'leaning toward statistical significance' may thus be clinically relevant for enhancing practice, particularly in studies with small sample sizes. [41]

### 3.2. Lower the alpha (α) level in certain cases

Related to how we view p-values is the common practice of lowering the conventional significance level of .05. A convincing justification for that practice is increasing the likelihood of a study's reproducibility. [42] Nevertheless, when 72 statisticians and researchers advocated changing the statistically significant p-value from p < .05 to p < .005 [43], the proposed adjustment revealed certain side effects, including a conflict with statistical power and the types of studies that can be conducted to achieve the sample sizes required. To achieve the same power of 80%, α = .005 requires a 70% larger sample size than α = .05, which could result in fewer smaller exploratory studies due to limited resources among researchers. [44] Although some might argue that studies with sufficiently large sample sizes should be conducted, not several small studies instead, novel topics and areas are constantly emerging that need to be investigated with smaller-scale exploratory research approaches. At the same time, adjusting p-values may be necessary in some cases, including in multiple comparison testing. For that reason, whether multiple comparison testing should be used warrants careful consideration and, if deemed appropriate, then so does the choice of test.

*Multiple comparison testing*
In multiple comparison testing, two rival traditions dominate: the classical (i.e., frequentist) view and the rational (i.e., epidemiological) view. [22] In the classical tradition, correcting for multiple tests is deemed necessary because reducing the rate of Type I errors is a priority. In that approach, the greater the number of comparisons, the greater the risk of false positives. The rational tradition, by contrast, argues that such correction defies common sense for two major reasons. On the one hand, p-value adjustments are calculated based on how many tests are conducted, and that number is arbitrary and variable. On the other hand, although p-value adjustments reduce the likelihood of making Type I errors, they increase the odds of making Type II errors (i.e., false negatives), or else needing to increase the sample size. [45] Following that logic, p-values should never be substituted for scientific reasoning, and even if a corrected p-value makes an association non-significant, the correction remains based on an assumed dichotomy between p-values above and below certain thresholds.

That said, I generally recommend adjusting for multiple comparisons to avoid rejecting the null hypothesis too hastily. Critics might counter that the theoretical basis for recommending a routine adjustment for multiple comparisons results in overemphasizing probability and significance testing. [46] However, supported by research examining common problematic statistical practices in nursing research [47,48], a more pressing problem is probably that nursing researchers tend to ignore the problems of multiple testing altogether.

Still, multiple comparison testing should not be used uncritically. Certain study-specific circumstances should be considered. Particularly, it is not always necessary to correct for multiple testing in the case of a small set of planned comparisons (i.e., scientifically sound comparisons) or in exploratory examinations aiming to generate hypotheses through post-hoc tests of unplanned comparisons.[49] Conversely, corrections are particularly useful in three situations: *a)* when it is critical to avoid a Type I error, *b)* when a single test of the null hypothesis that no results are significant is required, *c)* and when many tests are conducted without any hypotheses in place, as in entirely exploratory research. [49,50]

Because multiple comparison testing comes in many forms, choosing an appropriate one requires considering how well it will perform in balancing Type I and II error rates. To that purpose, two additional terms are often used: Type I error rate per comparison, which represents the probability of the incorrect rejection of the true null hypothesis when a comparison is tested, and family-wise Type I error rate, which represents the probability of a Type I error when a series (i.e., family) of tests is performed.

Multiple comparison tests can generally be categorized into three types according to their principles: *pairwise comparison tests*, *restricted sets of contrast tests*, and *post hoc error correction tests*. First, pairwise comparison tests compare all possible pairs of group-based means from k number of groups, such that the total number of independent comparisons is $k(k-1)/2$. Second, restricted sets of contrast tests are appropriate for a small number (e.g., <10) of tests; otherwise, their results are somewhat conservative when applied for many tests. Last, post hoc error correction tests are conducted as post hoc error corrections after all planned comparisons have been assessed. Such tests explore all possible complex relationships and are applied with the most stringent error control.

Beyond that, it is pivotal to know whether the data come from and/or exhibit parametric or non-parametric properties, which should be considered when choosing the type of multiple comparison test to perform. Incorporating those considerations [41,51,52], Table 1 details some of the most common multiple comparison tests.

**Table 1.** Overview of some of the most used multiple comparison tests.

| Principle | Test | Use | Notes |
|---|---|---|---|
| Pairwise comparison | Mann–Whitney U test | NP | *When:* If an omnibus test, such as the Kruskal–Wallis H test (KW), yields significance, it is used to identify differences among independent groups. *Alternative:* Dunn's test, even for unplanned comparisons (i.e., exploratory post hoc analyses). *Strengths:* It can be employed when the data is skewed since ranking relaxes the influence of extreme values. *Limitations:* Inappropriate for pairwise comparisons when KW is rejected because the rank sum test and KW employ different rank orderings; in such cases, Dunn's test should be used. |
| | Dunn's test | NP | *When:* If an *omnibus* test (e.g., KW) is rejected. *Alternative:* The Nemenyi joint rank test when ranking order matters. *Strengths:* Preserves rankings instead of re-ranking them (e.g., Mann–Whitney) when comparing treatments, allowing control of $\alpha$FW through Dunn's proposed Bonferroni adjustment. *Limitations:* Can be less powerful and occasionally overly conservative. |
| | Tukey's HSD (or simply, 'Tukey') | P | *When:* After a statistically significant analysis of variance (ANOVA) identifies where differences exist. *Alternative:* The Tukey–Kramer test, preferred for unequal group sizes. *Strengths:* The simplest method to control $\alpha$FW, and preferable when conducting all pairwise comparisons. *Limitations:* Assumes homogeneity of variance across groups. |
| Restricted sets of contrasts | Bonferroni | P | *When:* To control experiment-wise error rates after multiple t-tests or following ANOVA for $\alpha$FW correction. *Alternative:* Sequential Bonferroni tests (Holm–Bonferroni method). *Strengths:* Intuitive and straightforward calculation. *Limitations:* Can be overly conservative and less powerful with a large set of tests. |
| | Holm–Bonferroni | P | *When:* To correct $\alpha$FW at $\alpha$, but with reduced Type II error risk compared to the Bonferroni method. *Alternative:* The Šidák test, a less conservative yet stable approach. *Strengths:* Corrects Type I errors as effectively as the Bonferroni method while retaining more statistical power; suitable for $\alpha$FW adjustment. *Limitations:* Lower power with numerous tests. |
| Post hoc error correction | Dunnett's test | P | *When:* To compare means from multiple experimental groups to one control group. *Alternative:* Dunnett's C test when equal variance assumption is violated. *Strengths:* Supports one- and two-tailed testing. *Limitations:* Compares groups only to the control group, not among themselves. |
| | Scheffé's | P | *When:* To adjust $\alpha$ in linear regression analysis; applies to the set of estimates for all possible contrasts among the factor level means, not just pairwise differences. *Alternative:* The Tukey–Kramer test for a limited number of comparisons. *Strengths:* Offers flexibility by allowing the analysis of all linear contrasts and strict error control; suitable for unequal sample sizes between groups. *Limitations:* May be overly conservative and less powerful with small samples. |

Note. NP = nonparametric, P = parametric, $\alpha$FW = familywise $\alpha$ error level.

Regarding the recommended test, for non-parametric post hoc pairwise comparisons, Dunn's test is usually preferred because it preserves the ranking of the original data. However, for non-parametric post hoc pairwise comparisons, Tukey's honestly significant difference procedure is more commonly recommended due to its wide usage and solid performance under various conditions. In contrast, when comparing the means of a small number of groups or conducting preplanned comparisons among selected groups, the Bonferroni method, or ideally the Holm–Bonferroni procedure, is recommended. Furthermore, when comparing a control group to other experimental groups, Dunnett's test may be the best choice, whereas if you have a broad range of complex tests to consider, Scheffé's method is appropriate. [51,53]

Of course, other approaches for multiple comparison testing not elaborated herein are also available. [53] One increasingly employed group of analyses is permutation tests (e.g., exact tests), which are particularly useful when data are not normally distributed and when sample sizes are small. Although some limitations are that they can be computationally intensive and time-consuming, their increasing inclusion in common statistical software programs partly mitigates those potential setbacks. [29,54]

## 4. Communicating statistical significance

The discourse surrounding statistics—that is, how statistical information is communicated in writing and in speech—is a key aspect to consider when seeking to enhance nursing education and research practices involving the use of statistical significance.

### 4.1. Use correct terminology.

As noted, what p-values conceptually represent is often poorly understood, unlike seemingly more intuitive indications of statistical evidence based on likelihood theory. [55] Such misunderstanding is probably a reason why statistical significance and p-values are miscommunicated. It also highlights the need for nursing scholars to place particular focus on using correct terminology when communicating about those topics, whether during teaching or when writing research articles. To quote Greenland [35(p107)], "Several criticisms of p-values [...] are instead problems of poor teaching and terminology, and several properties of p-values [...] are often presented as fatal flaws but reflect instead how valid p-values should behave".

One such inaccurate use of terminology is suggesting that p-values normally exaggerate the evidence against the null hypothesis or that p-values often provide more evidence against the null hypothesis than is the case. [17] P-values inform us about the degree of agreement between a specific interpretation of the data (e.g., no relationship between variables) and the data itself. It does not provide information about the likelihood of this interpretation.[1] Generally, statistics involve describing quantities from distributions and related uncertainty and variability. Thus, statistical testing is not inherently confirmatory although it can provide a level of confidence about estimates. [53] A recommendation for nurse–educators and nurse–researchers is therefore to avoid stating that certain tests "confirm" their idea or notion, and to instead state how much of the uncertainty that has been reduced.

A similar misunderstanding is that p-values represent the probability that the null hypothesis is true, which is inaccurate because hypotheses lack probability in classical statistics. [22,56] Moreover, p-values do not indicate the likelihood of committing the Type I error of falsely rejecting null hypotheses. Another common misconception is that p-values indicate the size or importance of effects. However, unlike effect size measures, small p-values (e.g., .003) do not directly translate to strong associations, as discussed later. For that reason, referring to p-values when discussing the strength of associations is not recommended.

Another way in which statistical information can be miscommunicated relates to a general criticism of testing statistical significance, namely that it may lead users to focus on irrelevant null hypotheses, many of which are nevertheless irrelevant from a scientific perspective. [57] However, that problem is not a fault of p-values but the result of traditional training that encourages users to focus on such hypotheses. Despite repeated calls to abolish misleading jargon involving the word significance [39], few efforts have been made to correct Fisher's flaw of using null hypotheses for any tested hypothesis, thereby disregarding that, in plain English, null is a synonym for zero. That convention has led many users to assume that statistical testing involves testing null hypotheses only (e.g., without any association or effect) instead of all relevant hypotheses. [58]

### 4.2. Do not confuse statistical with clinical significance or importance.
A highly important source of confusion among nurse–educators, nurse–researchers, and practicing nurses is the difference between statistical and clinical significance. Edgeworth's [59] original intention for using statistical significance, articulated in 1885 and widely disseminated by Fisher [37] beginning in 1925, was foremost to indicate when a finding warranted further scrutiny. Statistical significance was thus not meant to suggest scientific importance [60].

In isolation, significance should be used to denote the substantive importance of research, whereas statistical significance is the likelihood of results due to chance. [61] For that reason, statistical significance should be differentiated from clinical significance. Although the difference may seem obvious and, in turn, widely known and established, research shows that the issue remains common in studies in health-related fields. [21]

Clinical significance is usually determined based on a particular health care intervention's practical value or relevance and may or may not consider statistical significance as an initial criterion. [61] When interpreting findings reported as being (statistically) significant or not, nurses should appraise the research's method and results critically and consider its clinical significance as well as importance for patients. [62,63] Some nursing scholars have more clearly integrated patients' perspectives into the concept of clinical significance and thus interpreted it as the impact on the patient's life from the patient's perspective. [64] By contrast, its practical significance concerns the magnitude of effects as interpreted by researchers and/or clinicians.

For the reasons mentioned above, some scholars propose using the term 'importance' instead of 'significance' when discussing effects altogether. [20] While this approach appears reasonable, I believe it is equally crucial to expand the understanding of clinical significance/importance by clearly considering it based on the perspectives of all relevant parties, including patients, their close ones, clinicians, and researchers. It is important to acknowledge patients' viewpoints, which may diverge from those of clinicians. To evaluate clinical significance from the patients' perspective, the use of validated surveys that encompass group-level indicators (e.g., effect size indices) and individual-level benchmarks (e.g., the minimal important change index) has been recommended.[19]

## 5. Using additional key metrics
As can be imagined, nursing scholars should consistently use metrics that convey the meaning and level of uncertainty of their research findings better than p-values. To achieve this goal, reporting confidence intervals (CIs) and effect size measures should be mandatory. CIs offer a superior measure to p-values for assessing the precision of a point estimate (i.e., using sample data to approximate a population parameter, such as the population mean). Effect size measures indicate the strength of the relationship between variables and, in contrast to p-values and CIs, provide insights into clinical significance.

### 5.1. Always report confidence intervals and effect sizes

Simply put, a CI represents the range of values in a sample that are likely to contain the population parameter if the test is repeated in a similar fashion over the long term (provided that biases, such as sampling bias, are not present). Confidence is usually set at 1 minus the level of significance, α, used in the statistical test. Thus, if α is set at p < .05, then the CI would be 1 − .05 = .95—that is, 95%. Accordingly, if the test were repeated several times, then 95% of the CIs can be expected to capture the population parameter (e.g., population mean).

CIs can be calculated for many different tests and point estimates, including the t-test, chi-square test, ANOVA, and regression analyses. [63]

Although calculations of CI and p-values are based on the same mathematical framework [4], focusing on interpreting data with CIs instead of p-values offers several advantages. [61,63] *First*, CIs provide a clinically easy-to-understand range estimate because the values are at the level of data measurement—for example, "mm/Hg" for blood pressure comparisons. *Second*, CIs provide information about how precise point estimates are. Given an equally large sample size and sample variability, the CI of 95% would be quite wide for an uncertain estimate but narrower for a more certain estimate. *Third*, statistically significant results can be assessed using CIs. For example, if the CI of 95% for the difference between two groups includes a value of 0, then the p-value will exceed .05. Similarly, when comparing groups using a ratio (e.g., odds ratio), no statistically significant difference between groups is indicated if the value of 1 is encompassed within the CI. *Fourth* and finally, CIs can be used in meta-analyses and in meta-analytic reasoning, which considers effects across studies instead of focusing on single studies. CIs permit researchers to compare their results more directly with CIs from past research and are useful in random-effects meta-analyses because they quantify uncertainty in the point estimates of interest.

Although CIs contain more useful information than p-values, they do not provide information about how much of an effect a nursing intervention achieved. Thus, effect size measurements are needed. Effect sizes can be used to estimate the magnitude of an effect according to several different tests. Common tests and their often-used effect size measurements include correlations between two variables (e.g., Pearson's r), associations between two categorical variables (e.g., Cramer's phi), mean differences between two groups (e.g., Cohen's d), or the likelihood that an event will occur given a certain stimulus (e.g., odds ratio).

Effect sizes can be standardized (e.g., Cohen's d) or unstandardized (e.g., unstandardized beta coefficient in regression models). Standardized measures, as unit-less measures, are preferable when results from several studies are combined, as in meta-analyses. For similar reasons, they are also used when comparing the effects of variables measured in different units in multivariate analyses, as in multiple regressions. [65]

There are two primary ways of estimating the relative size of an effect size; the observed value can be compared with previously reported effects from similar studies, or it can be evaluated based on conventional threshold values for the specific effect size measure. [66] For example, Cohen defined the effect size measures for Cohen's d as 'small' (d ≤ .2), 'medium' (d ≥ .5), and 'large' (d ≥ .8). [57] However, like p-values, these thresholds are context-dependent and based on statistical distributions. [20] Like 'statistical significance,' they do not inherently imply clinical importance. A 'small' effect can indeed be significant if it relates to a critical health outcome that affects a substantial portion of the population, such as blood pressure. Ultimately, this underscores the point made in section 4.2: the importance of distinguishing between 'statistical' and 'clinical' significance.

For another advantage, unlike p-values, most types of observed effect sizes are not affected by sample sizes, meaning that an effect's size can be estimated regardless of the sample size. However, effect sizes are functionally linked with sample sizes given their use to estimate adequate sample sizes for testing statistical significance (i.e., power analysis). For power analysis, a frequent measure is Cohen's d, lower values of which indicate that a larger sample size is needed, and vice versa. [65] An adequate sample size can subsequently be determined together with additional parameters of the selected significance level (α) and statistical power.

Taken together, effect sizes, preferably standardized, should be reported to allow users to indicate whether a clinically relevant effect was identified. If not, then it can be used in power analysis to determine the sample size required when planning future studies. [67,68]
Lastly, when discussing effect sizes and their implications, it is important to consider the study design, as the term "effect" implies a causal relationship between variables. But strictly speaking, in observational studies, the values are not "effects" but rather "associations". [20]

### 5.2. Consider the Bayes factor
Most tests, estimates, and approaches discussed in this article exist within the realm of frequentist inference, a pillar of classical statistics. Per that reasoning, only repeatable random events possess probabilities, which thus equal the long-term frequency of the occurrence of the events in question. By contrast, Bayesian inference regards probabilities as a more general concept, namely as representing the uncertainty of any event or hypothesis, even non-repeatable ones, as prior knowledge of what might influence whether the event is included in the estimation. [1,55,69,70]

First, when I suggest 'considering' the Bayes factor, it can encompass a few approaches: *a)* using and reporting both p-values and Bayes factors together, *b)* interpreting p-values with respect to their associated Bayes factor bound (BFB), or *c)* transitioning from frequentist statistics to Bayesian statistics. [69,71] Here, I will focus on the latter.

Although Bayesian statistics became relatively popular in the 1950s, the epistemological principles underlying its view on probability were formulated by Thomas Bayes in the mid-1700s and are thus far older than those of the classical frequentist view. [23] For a meaningful discussion of the Bayes factor, its underlying principle (i.e., Bayes's theorem) needs to be introduced. In non-technical terms, the theorem can be stated as:

Known information + Data = Total information.

Known information, often termed the "prior distribution," represents the researcher's knowledge before the test, whereas data, referred to as "likelihood," represents what is learned from the empirical information. Last, total information, or the "posterior distribution," thus considers the researcher's preexisting knowledge and the knowledge gained from the data. For hypothesis testing, that formula is compared with a classical analysis that excludes known information (i.e., prior distribution), stated as Data = Total information.

Clearly, if there is no known information of the phenomenon under study (prior distribution = 0), then both formulas are the same. Bayesian inference is therefore theoretically akin to classical analysis under such circumstances. Following Bayes's theorem, the Bayes factor constitutes the ratio of the probability of the observed data, conditional on two competing hypotheses. As such, it can be interpreted as a measure of the strength of evidence in favor of one hypothesis over another. The Bayes factor can also be used to quantify evidence in the data for null and alternative hypotheses. [72]

In nursing research, the Bayes factor offers several advantages. [71,73,74] First, it does not require the dichotomous decision to reject or not reject null hypotheses. Instead, it offers evidence in favor of each hypothesis under consideration and can be used to quantify data in favor of the null hypothesis. It can also be used to evaluate multiple hypotheses while considering that multiple hypotheses are evaluated, not a single rival hypothesis. During data collection, support for the hypotheses of interest can be continually updated (i.e., Bayesian updating); after all, if data-based support for the hypotheses of interest is unconvincing as a study is being conducted, it is acceptable within the Bayesian paradigm to collect more data and reevaluate the hypotheses.

Bayes factors provide a coherent approach to determining whether non-significant results support a null hypothesis over a theory or whether the data are simply insensitive. More specifically, Bayes factors use the data themselves to determine their sensitivity in distinguishing theories, as well as use aspects of a theory's predictions that are often easiest to specify. This is advantageous compared to the p-value. A high p-value cannot independently count against a theory that predicted a difference, since a high p-value might be the result of data insensitivity due to for example, a high standard error. [42,71]

Other issues with Bayes factors merit mention, including subjectivity. Strictly speaking, Bayes factors do not control Type I and Type II errors but do control Bayesian error probabilities. To assess error probabilities, researchers first have to define the prior distribution, and their potentially subjective selection of what is prior therefore impacts the outcomes of their analyses.[72] A commonly proposed solution is to use default prior distributions, in which the prior parameters are fixed and independent of the problem or phenomenon under investigation. Albeit seemingly appealing, uncritical adherence to default values can introduce the same problem as default α values for significance when assessing p-values. In response, other solutions have been suggested, including defining priors more precisely via elicitation (e.g., information from previous applicable research) and conducting sensitivity analyses to evaluate the impact of prior assumptions. [75]

Last, a primary challenge in encouraging nursing scholars to use Bayesian factors is that researchers and students in nursing have almost exclusively been taught classical statistics and the frequentist perspective on probabilities. Changing such views on statistics can prove difficult, as can overcoming practical challenges such as varied possibilities in common statistical packages. Even so, the advantages of the approach cannot be ignored. For nurses, they include the ability to incorporate one's prior knowledge during hypothesis testing, including knowledge derived from clinical experience. Research has also suggested the feasibility of teaching basic Bayesian principles to nursing students [74], which, to be implemented successfully, should be gradually introduced (e.g., in short seminars in an otherwise frequentist-based statistics course), along with real-world examples in nursing. [74]

## 6. Aspects of organizational and data management

Aside from the computational and interpretational aspects related to statistical significance and p-values, other more overarching aspects pertaining to planning nursing research studies warrant consideration.

### 6.1. Preregister studies when relevant.

One of the most important and affordable improvements for obtaining reliable measures of statistical significance is to pre-register key elements of a study design. In that process, researchers outline such elements (e.g., sample size) and analytical procedures (e.g., number and sequence of tests) known to affect p-values before data collection. [41] Research has shown that preregistering the primary hypotheses of clinical studies decreased the proportion of statistically significant positive findings from 57% to 8%. [76] Considerable differences in effect sizes have also been identified between studies with and without preregistration. [66]

More preregistered studies are therefore needed to obtain reliable population effects. Ideally, journals should decide if research is important enough for publication based on the preregistration information and regardless of outcomes stated in p-values. Doing so would not only encourage pre-registration, because it would improve the likelihood of publication, but also help to counter effects of publication bias because more non-significant findings could be reported. Preregistration is easily accomplished—for example, with the Open Science Foundation. Researchers skeptical of preregistration should know that such repositories usually enable researchers to adjust the level of publicity if they are worried that other researchers might plagiarize their ideas.

### 6.2. Enable data availability

Another aspect related to the transparency of research is data sharing. In a perfect world, all raw data would be published to allow the assessment of a study's analysis, quality of data, and overall findings. When data are published in a repository, they should be accompanied by a permanent identifier (e.g., digital object identifier), time-stamped, and published in an unalterable format. [77] Data should be published in a raw, unprocessed form; otherwise, data might be biased due to undocumented data management practices, for seemingly innocuous alterations (e.g., variable coding) can considerably impact statistical outcomes.

The scientific push toward open science has made progress with the increase of data sharing policies adopted by research journals and data repositories. Although such policies are feasible with small volumes of data, they can be prohibitively expensive for highly complex data sets. [77] Even though that problem is not common in nursing research, researchers in nursing should encourage funding bodies to cover those potential costs for the sake of transparency and replicability. After all, it is in their interest to be associated with high-quality research.

Other approaches to increasing data sharing have focused on improving researchers' willingness to share their raw data. Some common incentives are different badge systems (e.g., open data badges) or the recognition of researchers who publish open data sets by creating a special category of research data authorship. [77] However, the effects of such incentives have been shown to be inconsistent. A complementary or alternative approach to publishing raw data is to publish data summaries or all sample analysis scripts used to generate all figures and tables. Albeit less complete and more sensitive to bias, doing so would preclude the problem of needing massive storage spaces.

## 7. Conclusions and implications

To be sure, clinical significance and statistical significance are different concepts. A study with statistically significant differences and a large sample size may be of less interest to clinicians than one with a smaller sample size, statistically non-significant results, and substantial measures of effect size. Nurses working in evidence-based practice should evaluate the clinical importance of the results of research by carefully evaluating the study's design, sample size, statistical power, likelihood of Type I and Type II errors, data analysis, and the presentation of statistical findings. [21]

Moreover, we should move away from the traditional emphasis on probability and statistical significance, to ensure that metrics such as effect sizes form the basis of teaching nursing students the concepts involved in inferential statistics.[68] It is also clear that p-values possess practical utility in offering insights into what is observed and stand as the first line of defense against being misled by randomness. Generally, most people would be more suspicious if a coin tossed was heads 19 times in a row than only five times in a row. Likewise, scholars are generally interested in how strong the evidence is against a null hypothesis (e.g., a p-value of .049 or .003).

No single index should or can substitute for scientific reasoning, however. As stated by a panel of nurse–statisticians, it is important for nursing scholars to realize that statistics is a discipline, not merely a set of tools. [78] Thus, imparting a conceptual, instead of procedural, understanding of statistics might facilitate nursing students' long-term understanding of statistics. [79] Interventions and educational elements shown to improve nurses and nursing students' statistical literacy include the use of real-life examples, visual teaching aids, group work, and journal clubs using research articles. [80–82] In fact, many of those methods of describing statistical findings to a non-specialist readership were used by Florence Nightingale, the prominent nurse and the first women admitted into the Royal Statistical Society. [83] Nightingale's way to conveying clinical findings by using methods of data visualization has been successful (e.g., the polar area diagram). [83,84]

Last, more education and training in Bayesian statistics is needed for both nursing scholars and students alike, for that way of viewing probability and approaching statistical testing is intuitive and affords many clinical advantages. [73]

81. Hagen B, Awosoga OA, Kellett P, Damgaard M. Fear and Loathing: Undergraduate Nursing Students' Experiences of a Mandatory Course in Applied Statistics. *International Journal of Nursing Education Scholarship*. 2013;10(1):27-34. doi:10.1515/ijnes-2012-0044

82. Schwartz TA. Flipping the Statistics Classroom in Nursing Education. *J Nurs Educ*. 2014;53(4):199-206. doi:10.3928/01484834-20140325-02

83. Kopf EW. Florence Nightingale as Statistician. *Publications of the American Statistical Association*. 1916;15(116):388. doi:10.2307/2965763

84. Spurlock D. Beyond *p* < .05: Toward a Nightingalean Perspective on Statistical Significance for Nursing Education Researchers. Spurlock DR, ed. *J Nurs Educ*. 2017;56(8):453-455. doi:10.3928/01484834-20170712-02
1